\documentclass[apex]{jjap3}
\usepackage{newtxtext,newtxmath}

\title{Fabrication of Superconducting YBa$_2$Cu$_4$O$_8$ Film via Coprecipitation}

\author{Hiroshi Hara$^{1,2}$, Shintaro Adachi$^1$\thanks{Corresponding author: ADACHI.Shintaro@nims.go.jp}, Ryo Matsumoto$^{1, 2}$,  Yoshito Saito$^{1, 2}$, Hiroyuki Takeya$^1$, and Yoshihiko Takano$^{1, 2}$}

\inst{$^1$MANA, National Institute for Materials Science (NIMS), 1-2-1 Sengen, Tsukuba, Ibaraki 305-0047, Japan \\
$^2$Graduate School of Pure and Applied Sciences, University of Tsukuba, 1-1-1 Tennodai, Tsukuba, Ibaraki 305-8577, Japan} 

\abst{We have successfully synthesized the c-axis oriented YBa$_2$Cu$_4$O$_8$ (Y-124) film on a SrTiO$_3$ (1 0 0) substrate via a coprecipitation and a dip-coating method. The precipitations including Y, Ba and Cu ions were obtained using their metal nitrates solution and the aqueous solution of sodium hydroxide and oxalic acid. Superconducting transition of the synthesized film was observed at 64-70 K corresponding to bulk Y-124 in the magnetization and resistance measurement. The developed method is suitable to synthesize superconducting Y-124 films for technological applications.}

\begin{document}
\maketitle

   The cuprate superconductors have two-dimensional superconducting properties such as the anisotropic critical current density\cite{Hikata1989, Chong1997, Nagao2001, Kato2004, Ayai2007, Suzuki2012, Tanaka2015}. This anisotropy is attributed to its structure alternately stacked with superconducting CuO$_2$ plane and insulating charge reservoir layer\cite{Karppinen1999, Keimer2015, Adachi2019}. Therefore, the grain orientation of the cuprate\cite{Foltyn2007, Fukushima2008, Horii2009} is required for practical use. Most research for high-temperature superconducting wire have been using rare-earth (RE)-based cuprate superconductor REBa$_2$Cu$_3$O$_{7-{\delta}}$ (RE-123) or Bi-based Bi$_2$Sr$_2$Ca$_2$Cu$_3$O$_{10+{\delta}}$\cite{Hikata1989, Kato2004, Ayai2007, Foltyn2007}. However, a thermal instability for oxygen contents in both RE-123 and Bi-2223 is one of the limiting factors in its potential application. On the other hand, REBa$_2$Cu$_4$O$_8$ (RE-124) has a rigorous thermal stability for oxygen contents at high temperature up to 800 $\degC$\cite{Karpinski1988, Karpinski1989, Morris1990, Wada1990}. In addition, the superconducting transition temperature ($\it T$$_c$) of RE-124 can be increased up to 90 K by a partial Ca substitution of RE site\cite{Miyatake1989, Funaki2012}. Hence RE-124 is an important material for superconducting application above liquid nitrogen temperature that does not contain toxic elements as well as RE-123 and Bi-2223. Up to now, synthesizing RE-124 superconductor to good quality for application has been considered to be difficult. Here we report the simple synthetic method of Y-124 films using a coprecipitation.

   The coprecipitation process in Y-124 was similar to the previous study\cite{Ho1991}. Starting materials of Y(NO$_3$)$_3$$\cdot$6H$_2$O, Ba(NO$_3$)$_2$ and Cu(NO$_3$)$_2$$\cdot$3H$_2$O with the cation ratio of Y:Ba:Cu=1:2:4 (using 0.0025 mol of Y$^{3+}$) were dissolved in distilled or ion-exchanged water. An aqueous solution of 0.02 mol of sodium hydroxide (NaOH) and 0.003 mol of oxalic acid were used as a precipitant. The aqueous solution of NaOH was firstly added into the metal nitrates solution, and subsequently the aqueous solution of oxalic acid was added into the residual solution, resulting in pale blue precipitations (see figure 1(a)). The precipitations were dispersed in ethanol and dip-coated on a SrTiO$_3$ (1 0 0) substrate (see figure 1(b)). The sample was preheated at 400 $\degC$ for 10-15 min and dip-coated again. The dip-coating and preheating processes were repeated several times. Finally, the sample was heated at 750 $\degC$ for 12-20 h in a tube furnace flowing oxygen gas. The schematic images of the above process are shown in figure 2(a-e).
   X-ray diffraction (XRD) pattern was measured by $\theta$-2$\theta$ method with a CuK$\alpha$ radiation using Mini Flex 600 (Rigaku). The sample morphology and the chemical composition were identified by a scanning electron microscope (SEM) using JSM6010-LA (JEOL) equipped with an energy dispersive X-ray analyzer (EDX). The temperature dependence of magnetization was measured using a magnetic property measurement system (MPMS; Quantum Design). The temperature dependence of resistance was measured by the standard four-probe method. Gold wires as four probes were attached on the film with silver paste (4922N; Du Pont Co., Ltd.).

    Figure 3 shows the XRD pattern of a synthesized Y-124 film with a SrTiO$_3$ (1 0 0) substrate. All diffraction peaks excluding the impurity phases of Y$_2$BaCuO$_5$ and CuO were indexed as (0 0 $\it l$) of Y124 phase only, indicating that a synthesized Y124 film was well oriented along c-axis.
    
    The SEM image of the synthesized Y-124 film reveals that the film exhibits a porous surface consisting aggregations of small particles with submicron order, as shown in Figures 4(a) and 4(b). The synthesized Y-124 film shows the rough surface rather than a smooth plate-like morphology often observed in superconducting cuprate films, which is a good agreement with the previous reports\cite{Kanamori1993, Manabe1991}. The cation ratio of the particles was estimated to be Y : Ba : Cu = 0.65 : 1.95 : 4.00 using the EDX analysis.
    
   Figure 5 shows the temperature dependence of magnetization for a synthesized Y124 film under field cooling (FC) and zero-field cooling (ZFC) with an external field of 10 Oe. The magnetization started to exhibit the large diamagnetic signal at 70 K according to superconductivity in the ZFC mode. In the FC mode, the magnetization rose and exhibited the positive values due to the magnetism of the SrTiO$_3$ substrate.

   Figure 6 shows the temperature dependence of resistance for a synthesized Y-124 film. The resistance started to drop at 80 K and reached zero at 64 K. The onset transition temperature ($\it T$$_c$$^{onset}$) corresponds to $\it T$$_c$ of a bulk Y124 ($\sim$80 K) and the zero resistance temperature ($\it T$$_c$$^{zero}$) is almost consistent with the $\it T$$_c$ estimated from the magnetization measurement. In the normal state (80 K $<$ T $<$ 300 K), the resistance showed metallic and upper-curvature shape coming from the nature of under-doped cuprate superconductor\cite{Ito1993, Watanabe1997, Ando2004, Barisic2013, Adachi2015, Proust2016, Hara2018}. It is assumed that such behavior appears universally in cuprates\cite{Ito1993, Watanabe1997, Ando2004, Barisic2013, Adachi2015, Proust2016, Hara2018}, due to the in-plane scattering rate of the electrons was reduced by the pseudogap opening\cite{Ito1993, Watanabe1997, Ando2004, Barisic2013, Adachi2015, Proust2016}. Near the surface of our Y-124 film was Y-poor composition suggested by EDX analysis, however, the temperature dependence was typical of in-plane characteristics of an as-grown RE-124 single crystal\cite{Proust2016, Hara2018}. Thus, we considered that the resistance in normal state was originated from the bulk properties of the synthesized Y-124 film.
   Our new method can produce the c-axis oriented and nearly single phase Y124 films by just dipping the precipitations on a substrate and heating. Although we suppose that further studies are required to increase the purity of Y-124 phase, we believe that the developed method is suitable to fabricate superconducting Y-124 films for various applications.

   In conclusion, the superconducting Y-124 films are successfully synthesized just by dipping the precipitations on a SrTiO$_3$ (1 0 0) substrate and heating. The precipitations are easily obtained using the Y, Ba and Cu nitrates solution and the aqueous precipitants of NaOH and oxalic acid. The fabricated film clearly exhibits superconductivity at 64-70 K in the temperature dependence of magnetization and resistance. Our Y-124 film was well oriented to a c-axis direction and showed in-plane characteristics. The developed method in this study is suitable to fabricate Y-124 films for superconducting applications.

\begin{figure}
\includegraphics{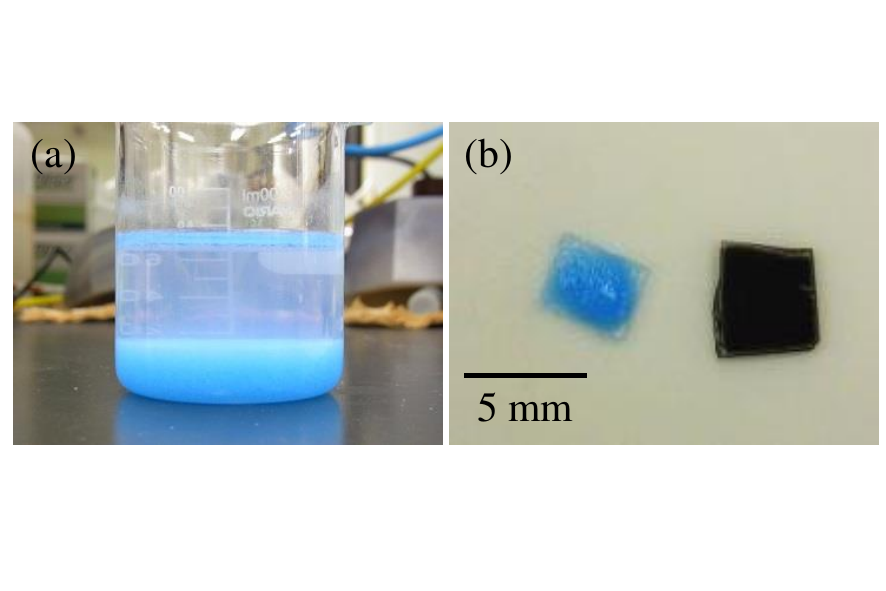}
\caption{(color online). Photographs (a) after coprecipitation, and (b) before (left) and after (right) heating the dip-coated samples.}
\label{f1}
\end{figure}

\begin{figure}
\includegraphics{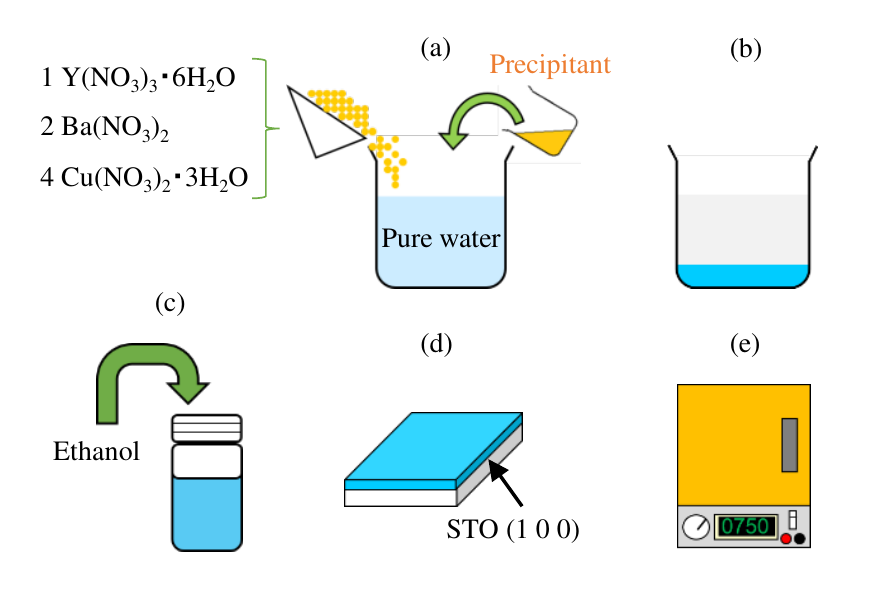}
\caption{(color online). Schematic image of the sample preparation process. (a, b) The coprecipitation process. (c) The dispersion of precipitations in the ethanol solution. (d) Dip-coating of dispersed precipitations on a SrTiO$_3$ (1 0 0) substrate. (e) Sample heated at 750 $\degC$ for 12-20 h in the furnace flowing the oxygen gas.}
\label{f2}
\end{figure}

\begin{figure}
\includegraphics{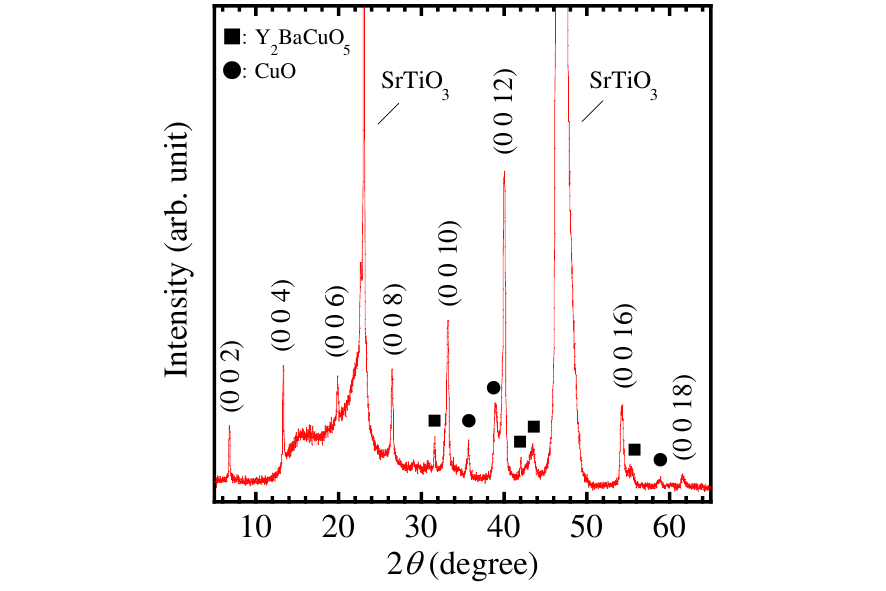}
\caption{(color online). XRD pattern of a synthesized Y124 film with a SrTiO$_3$ (1 0 0) substrate.}
\label{f3}
\end{figure}

\begin{figure}
\includegraphics{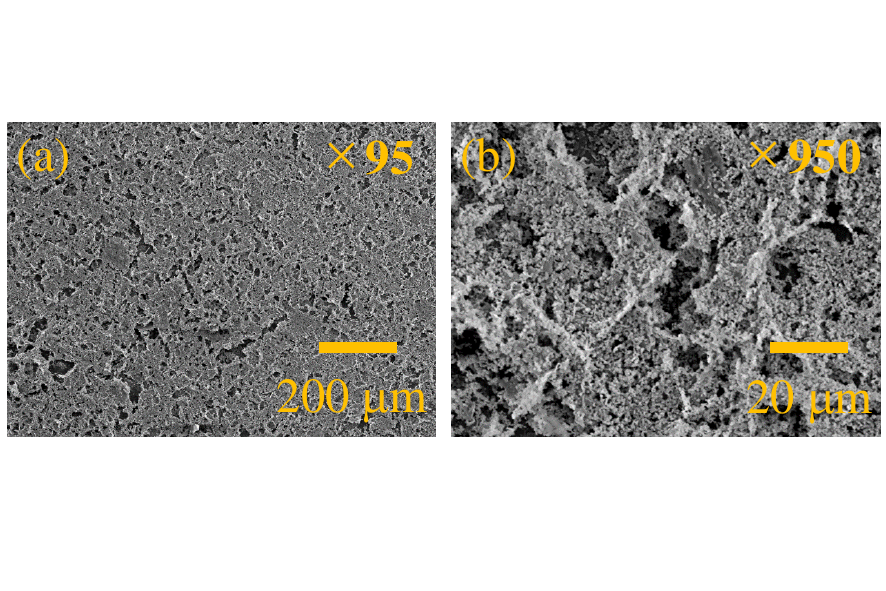}
\caption{(color online). SEM images of a surface of a synthesized Y124 film in (a) $\times$95 and (b) $\times$950 scales.}
\label{f4}
\end{figure}

\begin{figure}
\includegraphics{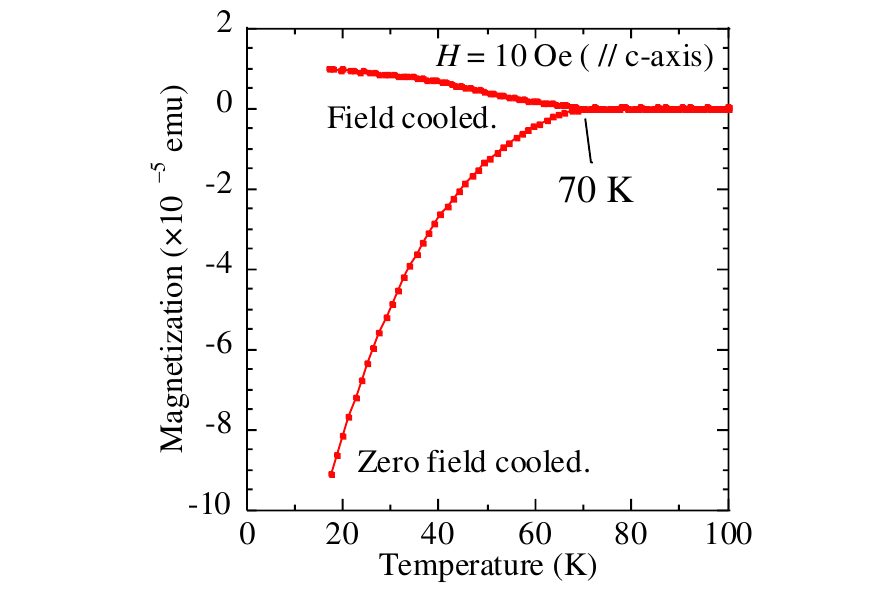}
\caption{(color online). Temperature dependence of magnetization for a synthesized Y124 film. Magnetic fields (= 10 Oe) were applied perpendicularly to the CuO$_2$ planes.}
\label{f5}
\end{figure}

\begin{figure}
\includegraphics{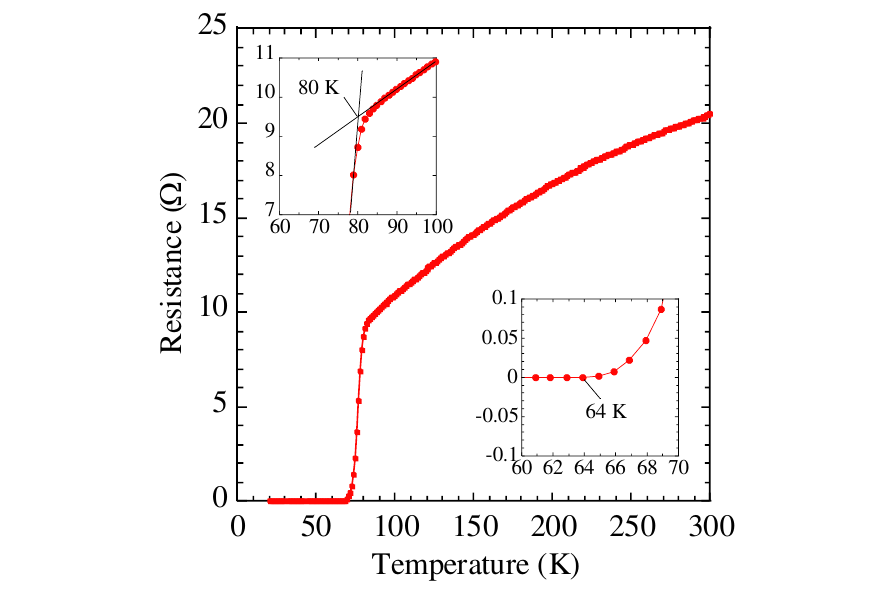}
\caption{(color online). Temperature dependence of resistance for a synthesized Y124 film. Left upper inset shows the enlargement view near the resistance drop, and right bottom one shows the enlargement view near the zero resistance.}
\label{f6}
\end{figure}

\begin{acknowledgment}
The authors greatly thank S. Harada, T. Ishiyama, T. Yamamoto and H. Okazaki for their support. This work was supported by JSPS KAKENHI Grant No. JP17J05926, JST-Mirai Program Grant No. JPMJMI17A2, and JST CREST Grant No. JPMJCR16Q6.
\end{acknowledgment}

\end{document}